\documentclass{aip-cp}

\usepackage[numbers]{natbib}
\usepackage{rotating}
\usepackage{graphicx}
\usepackage{amssymb}

\usepackage{savesym}
\savesymbol{iint}
\savesymbol{iiint}
\savesymbol{iiiint}
\savesymbol{idotsint}
\usepackage{amsmath}
\restoresymbol{AMS}{iint}
\restoresymbol{AMS}{iiint}
\restoresymbol{AMS}{iiiint}
\restoresymbol{AMS}{idotsint}

\usepackage{upgreek}

\begin{document}

\title{Radon Mitigation for the SuperCDMS SNOLAB Dark Matter Experiment}

\author[aff1]{J. Street\corref{cor1}}
\author[aff2]{R. Bunker}
\author[aff1]{E.H. Miller}
\author[aff1]{R.W. Schnee}
\author[aff3]{S. Snyder}
\author[aff1]{J. So}

\affil[aff1]{Department of Physics, South Dakota School of Mines \& Technology, Rapid City, SD 57701 USA}
\affil[aff2]{Pacific Northwest National Laboratory, Richland, WA 99352 USA}
\affil[aff3]{University of Delaware, Newark, DE 19716 USA}

\corresp[cor1]{Corresponding author: joseph.street@mines.sdsmt.edu}

\maketitle

\begin{abstract}
A potential background for the SuperCDMS SNOLAB dark matter experiment is from radon daughters that have plated out onto detector surfaces. To reach desired backgrounds, understanding plate-out rates during detector fabrication as well as mitigating radon in surrounding air is critical. A radon mitigated cleanroom planned at SNOLAB builds upon a system commissioned at the South Dakota School of Mines \& Technology (SD Mines). The ultra-low radon cleanroom at SD Mines has air supplied by a vacuum-swing-adsorption radon mitigation system that has achieved $>$1000$\times$ reduction for a cleanroom activity consistent with zero and $<$0.067\,Bq\,m$^{-3}$ at 90\% confidence. Our simulation of this system, validated against calibration data, provides opportunity for increased understanding and optimization for this and future systems.
\end{abstract}

\section{RADON BACKGROUNDS FOR SUPERCDMS SNOLAB}
A potential source of dominant backgrounds for many rare-event searches or screening detectors is from radon daughter plate-out~\cite{Simgen:2013dlh,Schnee:2014eea}. Backgrounds from $^{210}$Pb and the recoiling $^{206}$Pb nucleus from the $\alpha$ decay of $^{210}$Po were the dominant low-energy backgrounds for XMASS~\cite{Kobayashi:2015lla,Hiraide:2015cba}, SuperCDMS Soudan~\cite{Agnese:2014aa}, and EDELWEISS~\cite{Navick:2013hgx}. These backgrounds remain the dominant surface backgrounds for EDELWEISS-III~\cite{Armengaud:2017rzu}. Mitigation of radon daughters on surfaces has also been critical for SuperNEMO~\cite{Mott:2013nzg} and CUORE~\cite{wangSUthesis}. Both neutrons from ($\alpha,n$) reactions and $^{206}$Pb  recoils are important for LZ~\cite{lux2014backgrounds}, XENON1T~\cite{Aprile:2017ilq}, and DarkSide~\cite{Agnes:2014bvk}.

\begin{figure}[h]
  	\includegraphics[width=0.80\textwidth]{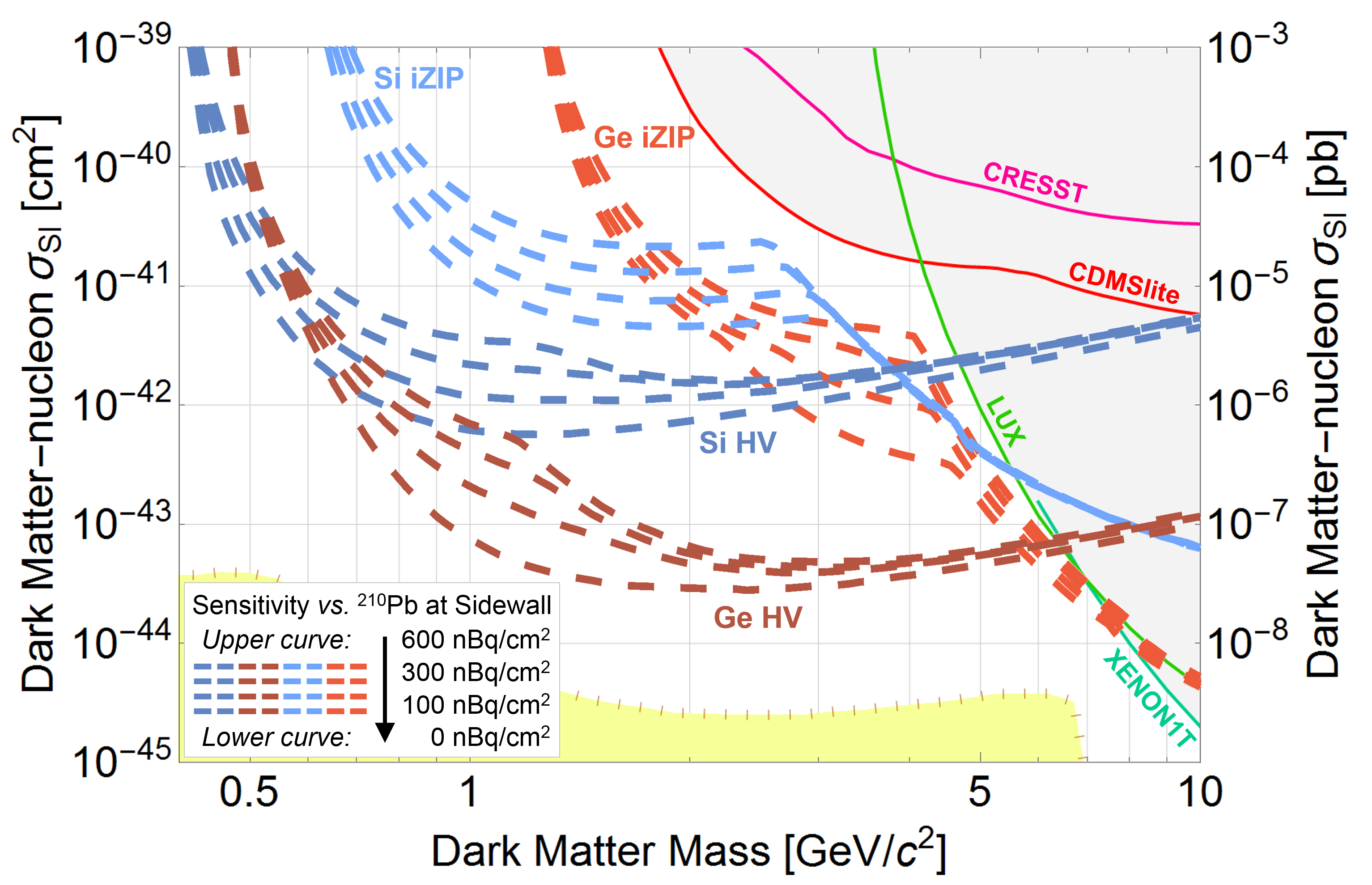}
  	\caption{Projected sensitivity to the spin-independent dark matter-nucleon scattering cross-section as a function of dark matter mass for Si and Ge iZIP and Si and Ge HV detectors of the SuperCDMS SNOLAB experiment~\cite{Agnese:2017jvy,Kurinsky:2016fvj}. Each set of four dashed lines represents projected sensitivities for (from bottom to top) 0, 100, 300, and 600\,nBq\,cm$^{-2}$ $^{210}$Pb surface contamination on the detector sidewalls (i.e., a sum across the detector and copper surfaces). At low masses, the sensitivity degrades significantly with increasing $^{210}$Pb contamination above 50\,nBq\,cm$^{-2}$.}
  	\label{sens_plot}
\end{figure}

Radon-daughter backgrounds are important to the expected low-mass sensitivity of the SuperCDMS SNOLAB experiment, which will use detectors of germanium and silicon to search for dark matter interactions~\cite{Agnese:2016cpb}.  Although the SuperCDMS Interleaved Z-sensitive Ionization and Phonon (iZIP) detectors provide excellent rejection of surface events above 8\,keV~\cite{Agnese:2013ixa}, at lower energies the rejection is expected to worsen to the point that radon-daughter backgrounds may dominate~\cite{Agnese:2016cpb}.  The situation is similar for the SuperCDMS high-voltage (HV) detectors, which provide the experiment's lowest energy threshold by amplifying the ionization signal~\cite{Agnese:2014aa,Agnese:2017jvy}.  Below energies of $\sim$0.5\,keV, rejection of events on the detector sidewall surface, based on the relative sizes of signals in the detectors' phonon sensors, becomes ineffective, so radon-daughter surface backgrounds are expected to dominate over bulk backgrounds for $^{210}$Pb concentrations $\gtrsim$50\,nBq\,cm$^{-2}$.  Figure~\ref{sens_plot} shows how the expected sensitivity of SuperCDMS SNOLAB varies for different surface concentrations of $^{210}$Pb.

Without mitigation, radon daughter plate-out during assembly underground at SNOLAB could dominate the expected background.  The duration of the SNOLAB assembly is estimated at about 150 hours.  At the average SNOLAB radon concentration of 130\,Bq\,m$^{-3}$, the expected plate-out would be about 70\,nBq\,cm$^{-2}$ of $^{210}$Pb.  To make this plate-out rate negligible, SuperCDMS SNOLAB has a goal of achieving a radon concentration underground of $<$0.1\,Bq\,m$^{-3}$ in its assembly cleanroom underground.

\section{THE DEMONSTRATION RADON-MITIGATION SYSTEM AT SD MINES}
Two systems of classes may create a radon-mitigated, breathable-air environment. One type flows continuously through a filtration column, while the second swings flow back and forth using two or more filtration columns. The filtration columns are usually filled with activated carbon. The continuous flow system (e.g.~\cite{Mount:2017aa,Nachab:2007zz}) operates on the basis that a considerable fraction of radon decays before exiting the column. For an ideal column, the final radon concentration $C_\text{final}=C_\text{initial}\exp{\left(-t_\mathrm{BT}/\tau_\mathrm{Rn}\right)}$, where $C_\text{initial}$ is the radon concentration of the input air, $t_\mathrm{BT}$ is the characteristic break-through time of the filter, and $\tau_\mathrm{Rn}=5.516$\,days is the mean lifetime of radon. The break-through time is the mean-time it takes for radon to pass through the carbon filter. To increase the break-through time, and therefore make a continuous flow system practical, one must cool the carbon to reduce desorption of radon. Continuous systems are commercially available (e.g.~\cite{Mount:2017aa,Nachab:2007zz}) and typically achieve reduction factors of $\sim$1000$\times$. 

\begin{figure}[!tbp]
	\begin{tabular}{c c}
  		\includegraphics[width=0.50\textwidth]{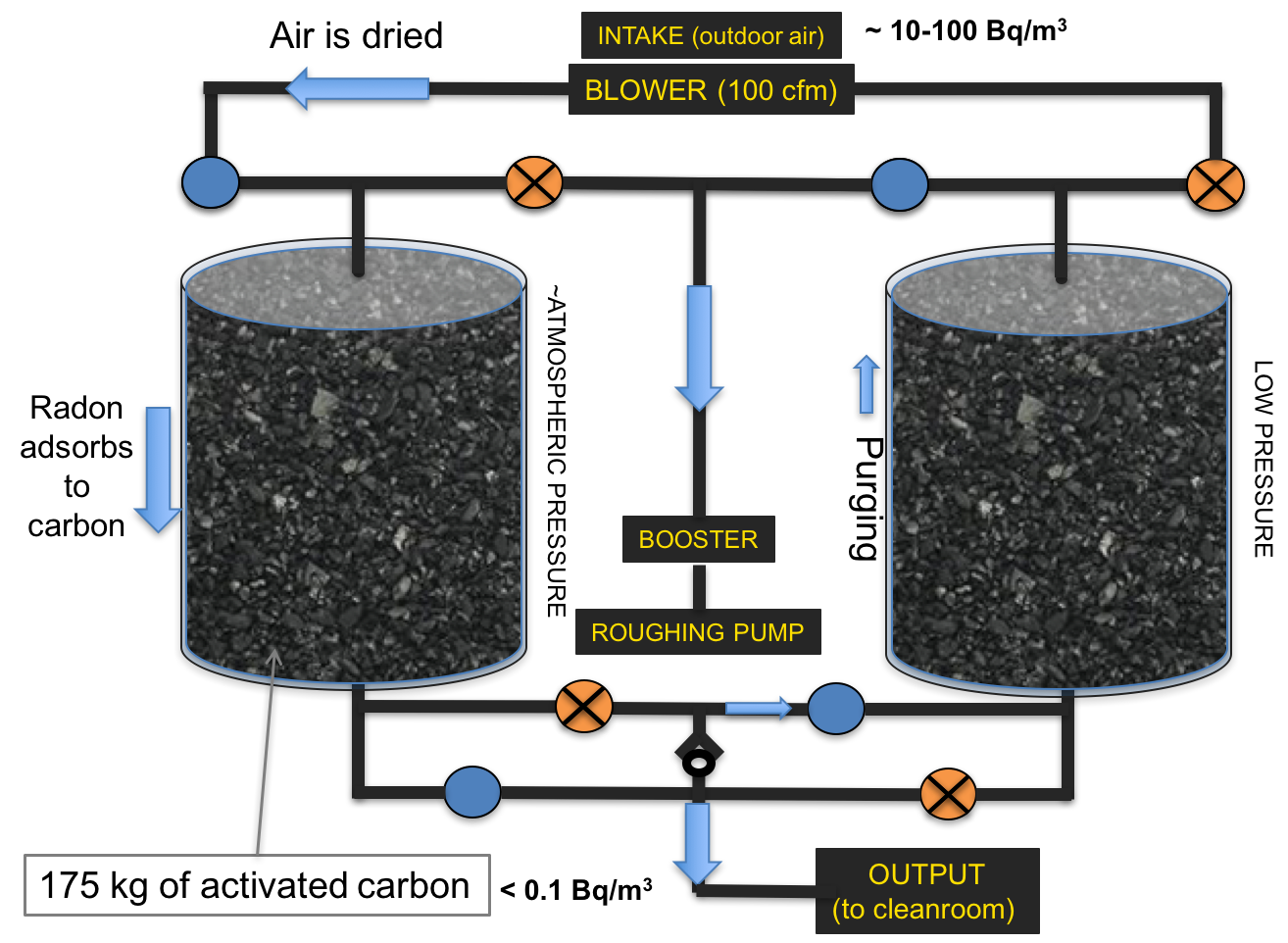} &
  		\includegraphics[width=0.46\textwidth]{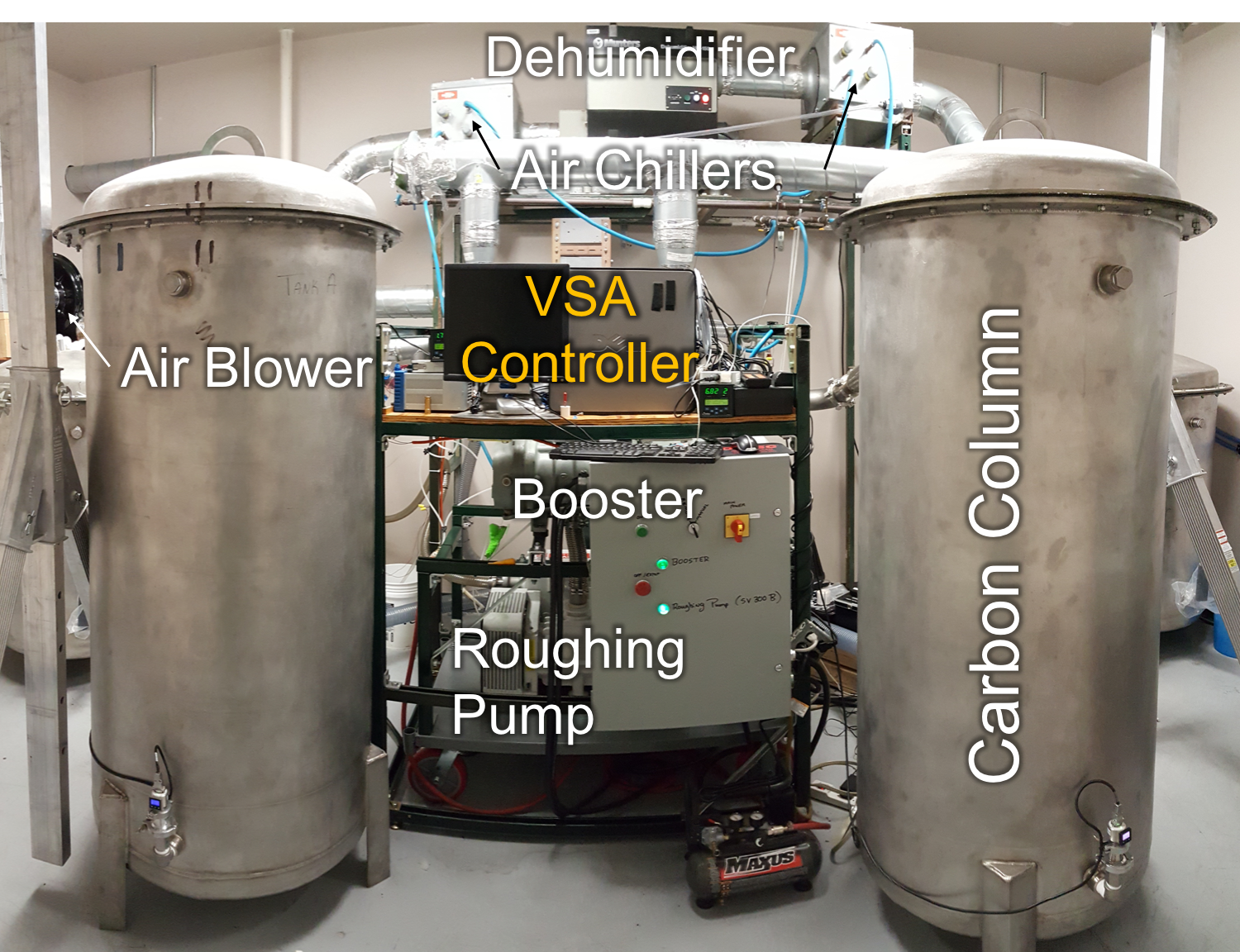}
  	\end{tabular}
  	\caption{\textit{Left}: Diagram of the vacuum-swing-adsorption radon mitigation system, showing the path of airflow (arrows) through open valves (solid circles), through the carbon tank at atmospheric pressure and to the system output, while ~10\% of the output air (small arrows) is diverted through a butterfly valve, backwards through the low-pressure carbon tank and exhausted at the vacuum pump. By opening the closed valves (crossed circles) and closing the opened ones, the roles of the tanks may be reversed. \textit{Right}: Image of the VSA system at SD Mines. Labels show the air blower, air chillers, dehumidifier, VSA system controller in front of valve cluster, carbon columns, roughing pump and booster.}
  	\label{VSA_Diagram_image}
\end{figure}

In a swing flow system, two or more filtration columns are used. While air is filtered through one column, the other is regenerated using either low pressure or high temperatures to allow radon to desorb efficiently and be exhausted. For a vacuum-swing-adsorption (VSA) system (e.g.~\cite{Pocar:2005kp,Pocar:2003dw,Street:2015nwa}), high-radon input air is filtered through the first column while the second column is pumped down to $\sim$20\,Torr. Well before the break-through time, the path of the high-radon input air is switched so that it flows through the second column instead, allowing the first to regenerate. For an ideal column, no radon reaches the output. Swing flow systems are more complicated both in operation and analysis. A VSA system (e.g. Fig.~\ref{VSA_Diagram_image}) can potentially outperform a continuous flow system even at a lower cost. Temperature-swing systems (e.g.~\cite{Hallin:LRT2010}) should provide best performance but at the highest cost and complexity.

The VSA system built at SD Mines (Fig.~\ref{VSA_Diagram_image}) was based closely on the Princeton design~\cite{Pocar:2003dw,Pocar:2005kp}. An air blower provides as much as $\sim$100 cubic feet per minute (cfm) to the VSA system input. The air is then chilled, dehumidified, and chilled again, to prevent moisture in the air from collecting on the carbon surfaces, and without heating the air (since radon desorbs from carbon more at lower temperatures). The high-radon air then passes through one of the two carbon columns. Radon adsorbs preferentially to carbon relative to nitrogen and oxygen because both radon and carbon are nonpolar and therefore experience induced dipole interactions described by the Van der Waals force. Because radon spends a higher fraction of its time adsorbed on the carbon, it moves with a net speed much slower than that of air. Before any radon has made it all the way through the column, the air flowing out is low-radon air. During this time (as much as hours), the second column is pumped down to $\sim$20\,Torr by a high-speed vacuum system and $\sim$10\% of the low-radon air flowing from the first column (called the purge flow) is introduced to the second column to help remove its radon. After less than an hour, the second column is regenerated (i.e., the radon is removed) and it may be used as a filter just as the first was. Before radon travels to the end of the first column, the system ``swings" such that high-radon air flows through the second column and the first is regenerated. In this way, low radon air is continuously supplied to the low-radon clean room without interruption.

An improvement to the original design of this system (e.g.~\cite{Pocar:2003dw,Street:2015nwa}) is in the way the regenerated column is brought back up to atmosphere. Previously~\cite{Pocar:2003dw,Street:2015nwa}, once a column was regenerated it was brought up to atmosphere by opening a second channel connected directly to the air blower, which bypassed the air chillers and dehumidifier. This choice maximized the amount of air supplied to the filling column in order to prevent air from being pulled from the clean room (which when under-pressured would pull high-radon air in from the lab environment). Our improved system instead closes the valve to the roughing pump after regeneration and then slowly fills the column up to atmosphere through the purge flow line; at 7\,cfm, it takes about 15 minutes to approach atmosphere (from $\sim$18\,Torr). This technique, termed the ``slow-fill method," reduces the number of valves needed by the VSA system, prevents the introduction of radon to the newly regenerated column before use, and maintains mechanical stability (because raising these columns to atmosphere quickly can produce sudden stresses on components).

\section{The Radon-Mitigated Clean Room and Results}
\begin{figure}[!tbp]
	\begin{tabular}{c c}
	\includegraphics[width=0.45\textwidth]{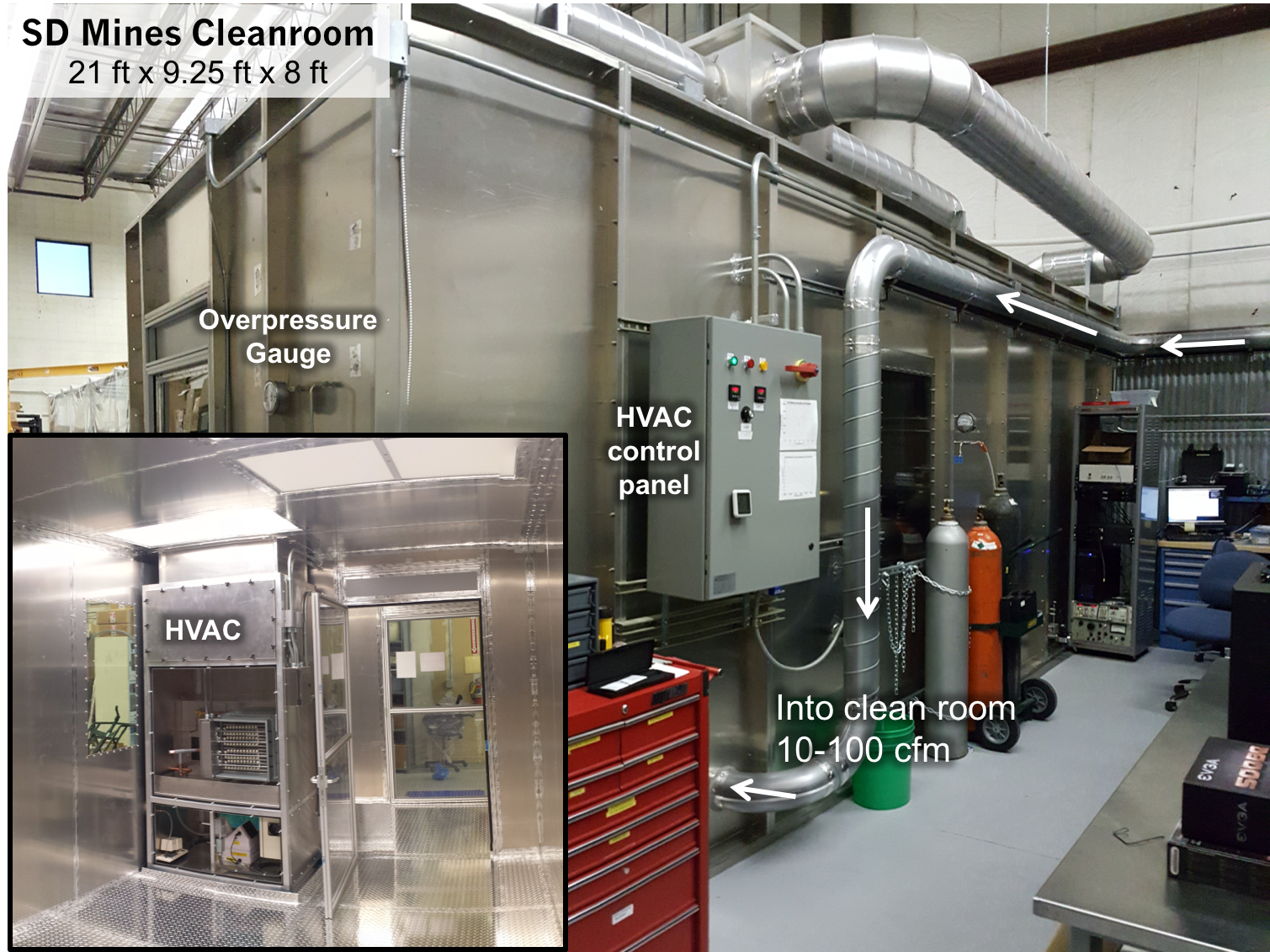} &
	\includegraphics[width=0.52\textwidth]{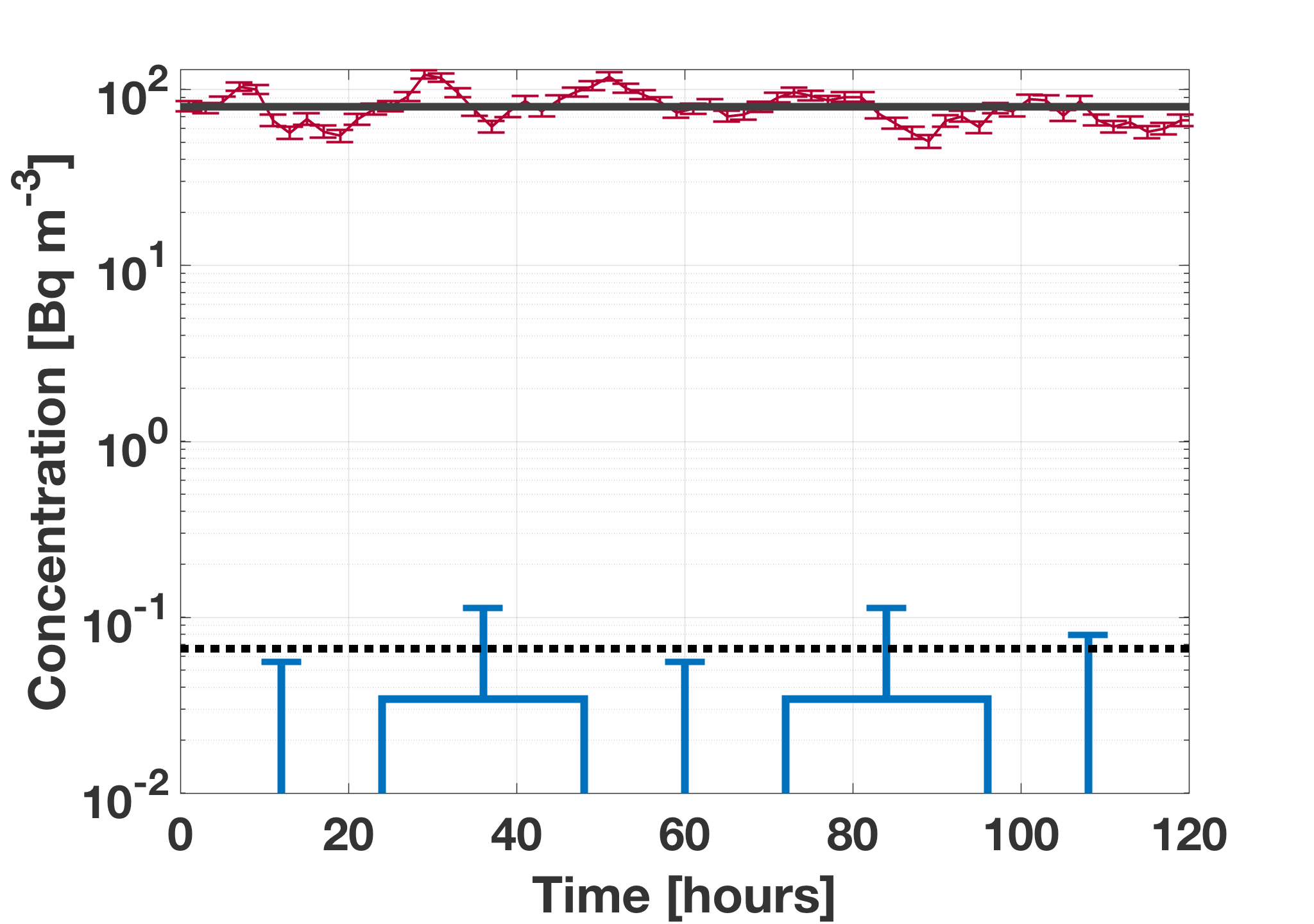}
	\end{tabular}
	\caption{\textit{Left}: The radon-mitigated clean room at SD Mines with interior view inset. Labels show an overpressure gauge, control panel for an internal heating, ventilation and air conditioning (HVAC) unit, the HVAC and steel ducting from the VSA system. The clean room is made from aluminum and has an HVAC that recirculates air ($\sim$1000\,cfm) through four HEPA filters. \textit{Left inset}: The HVAC inside avoids the difficulty of preventing high-radon air from the lab from being introduced to the clean room through the HVAC's recirculation blower region. \textit{Right}: Results of the SD Mines VSA system and clean room. The input air provided to the VSA system (upper error bars) had an average radon concentration (upper solid line) of $79.6\pm0.7\,$Bq$\,$m$^{-3}$, while the air inside the radon-mitigated clean room (dotted line) was measured to be $<$0.067\,Bq$\,$m$^{-3}$ at 90\% confidence. This shows a reduction of $>$1000$\times$. The solid curve with error bars represents the measured concentration in the clean room binned over 24 hours.}
	\label{CR}
\end{figure}

The VSA system supplies low-radon air for a radon-mitigated clean room has been commissioned at SD Mines (Fig.~\ref{CR}, \textit{Left}). The clean room was constructed primarily from aluminum, because radon does not easily diffuse through metals. The clean room also has polycarbonate windows that are sufficiently thick (6.4\,mm) such that radon decays before diffusing completely through. Make-up air is supplied at $10-90\,$cfm and replaces the clean room air volume quickly enough to make negligible radon emanating from materials inside the clean room. The achieved room radon concentration described below limits the activity from materials inside to be $<$1000\,Bq.

As shown in~\ref{CR} (\textit{Right}), the VSA system has demonstrated radon reduction meeting the goal of the SuperCDMS SNOLAB experiment. The upper limit on the radon concentration of the radon-mitigated clean room in February 2016 was $<$0.067\,Bq$\,$m$^{-3}$ at 90\% confidence. This shows a $>$1000$\times$ radon reduction between the VSA system input and the radon-mitigated clean room. Note that the VSA system output must have an even lower radon concentration than the cleanroom. This upper limit on radon concentration in the clean room is currently limited by the sensitivity of our Durridge Rad7 radon monitor used to make the measurement. 

\section{SIMULATION AND EXPECTED PERFORMANCE OF THE VSA SYSTEM}
The simulation of the VSA system provides predictions of performance outside of what we can directly measure. It aids in understanding our current VSA system and improvements of future systems in terms of cost and performance. The simulation models radon diffusion and translation through an activated carbon column, while accounting for column temperature, column pressure (as a function of the longitudinal distance in a column and time), radon emanation from the carbon (with activity of 0.01\,Bq\,kg$^{-1}$~\cite{Pocar:2003dw}) and radon decay. The simulation calculates the radon concentration as a function of position and time inside both of the VSA system's carbon columns during operation.

In each carbon column, the radon concentration (having units of Bq\,m$^{-3}$) may be denoted as the \textit{radon vector} $\boldsymbol{c}(x,t)$. Each element of the radon vector is the radon concentration in a position $x$ along the length of the column. Operators evolve these radon vectors in ways that represent the VSA system operation. There are three operators: one for filtering (flowing air forward through a column), one for regenerating a carbon column by flowing air backward through it, and one for slow-filling (bringing a column up to atmosphere). Each operator is a matrix in which the $j^\mathrm{th}$ row evolves the $j^\mathrm{th}$ position of $\boldsymbol{c}(x,t)$ by $\Delta t=1\,$min, which is the temporal resolution chosen for the simulation. The linear velocity of radon moving through a column is
\begin{equation}
	v_\mathrm{Rn}(x,t) = v_\mathrm{air}(x,t) A_\circ \exp{\left(-T_\circ/T\right)},
\end{equation}
where $v_\mathrm{air}(x,t)$ is the linear air flow at a position $x$ and time $t$ through the column, $A_\circ$ is the radon adsorption factor, and $T_\circ \approx 3500\,$K is the critical temperature of radon adsorbing to carbon~\cite{Strong:1979aa}. The operators are built upon a normal distribution that has been modified to account for a nonzero spatial binning and depends on the linear radon velocity $v_\mathrm{Rn}$ and the diffusion coefficient $D$ (which we assume is constant) of radon passing through carbon:
\begin{equation}
	\mathrm{Normal\_Dist}(x; v_\mathrm{Rn},D,j,\Delta t) = \frac{1}{2}\left[\mathrm{erf}\left(\dfrac{(x+\frac{1}{2}) -v_\mathrm{Rn}\Delta t - j}{\sqrt{4D\Delta t}}\right) - \mathrm{erf}\left(\dfrac{(x-\frac{1}{2}) -v_\mathrm{Rn}\Delta t - j}{\sqrt{4D\Delta t}}\right)\right] e^{-\Delta t/\tau},
	\label{dist}
\end{equation}
where $e^{-\Delta t/\tau}$ accounts for the decay of radon after one minutes and $\tau$ is the mean-lifetime of radon. This normal distribution model gives similar results to the n-theoretical stages framework (e.g.,~\cite{Pocar:2003dw,Strong:1979aa}). Each operator is built upon Eq.~\ref{dist}, but with different radon velocities to account for filtering, regeneration or slow-fill.

\subsection{Characterization and Expected Radon Concentration at the VSA System Output}
\begin{figure}[t]
	\begin{tabular}{c c}
	\includegraphics[width=0.5\textwidth]{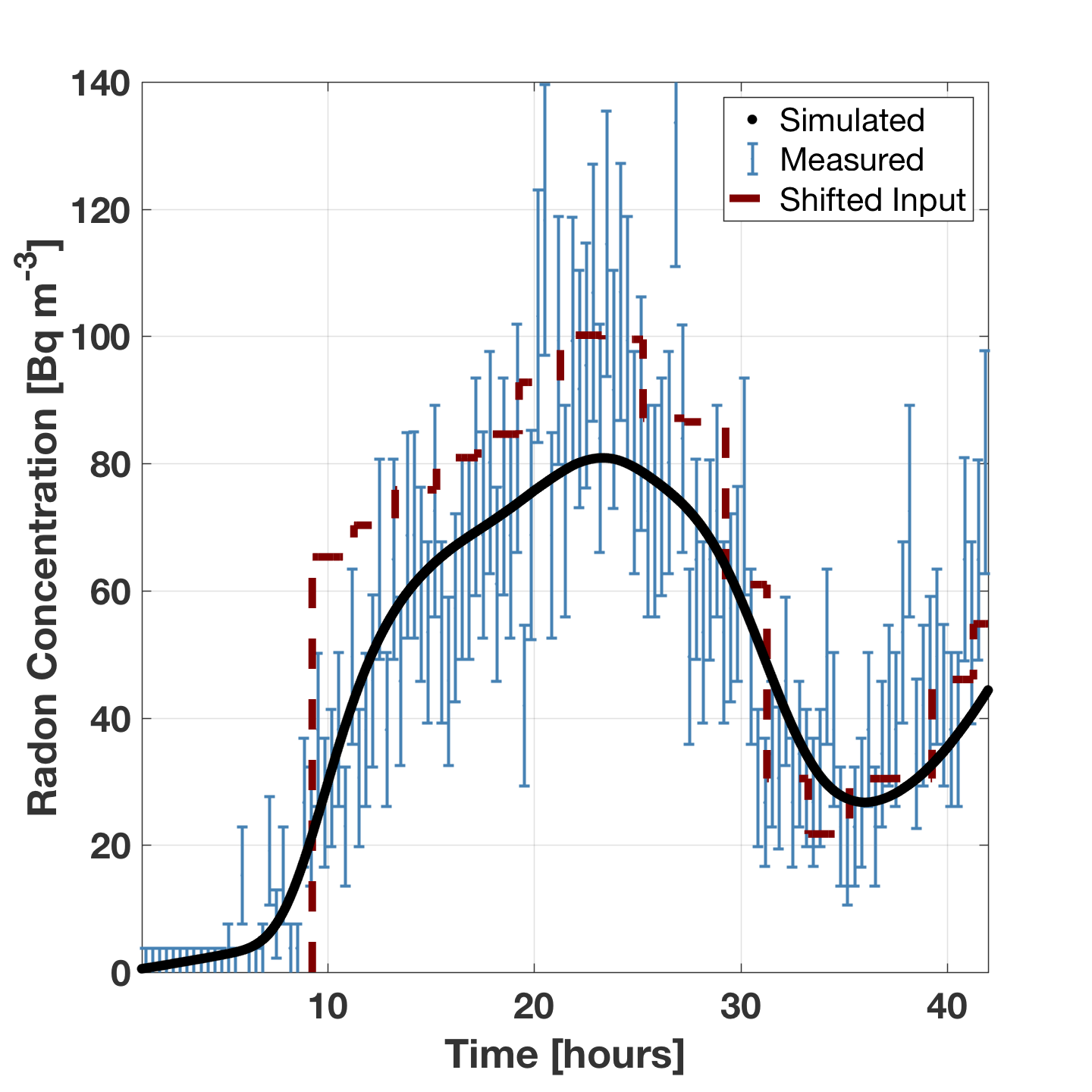} &
	\includegraphics[width=0.5\textwidth]{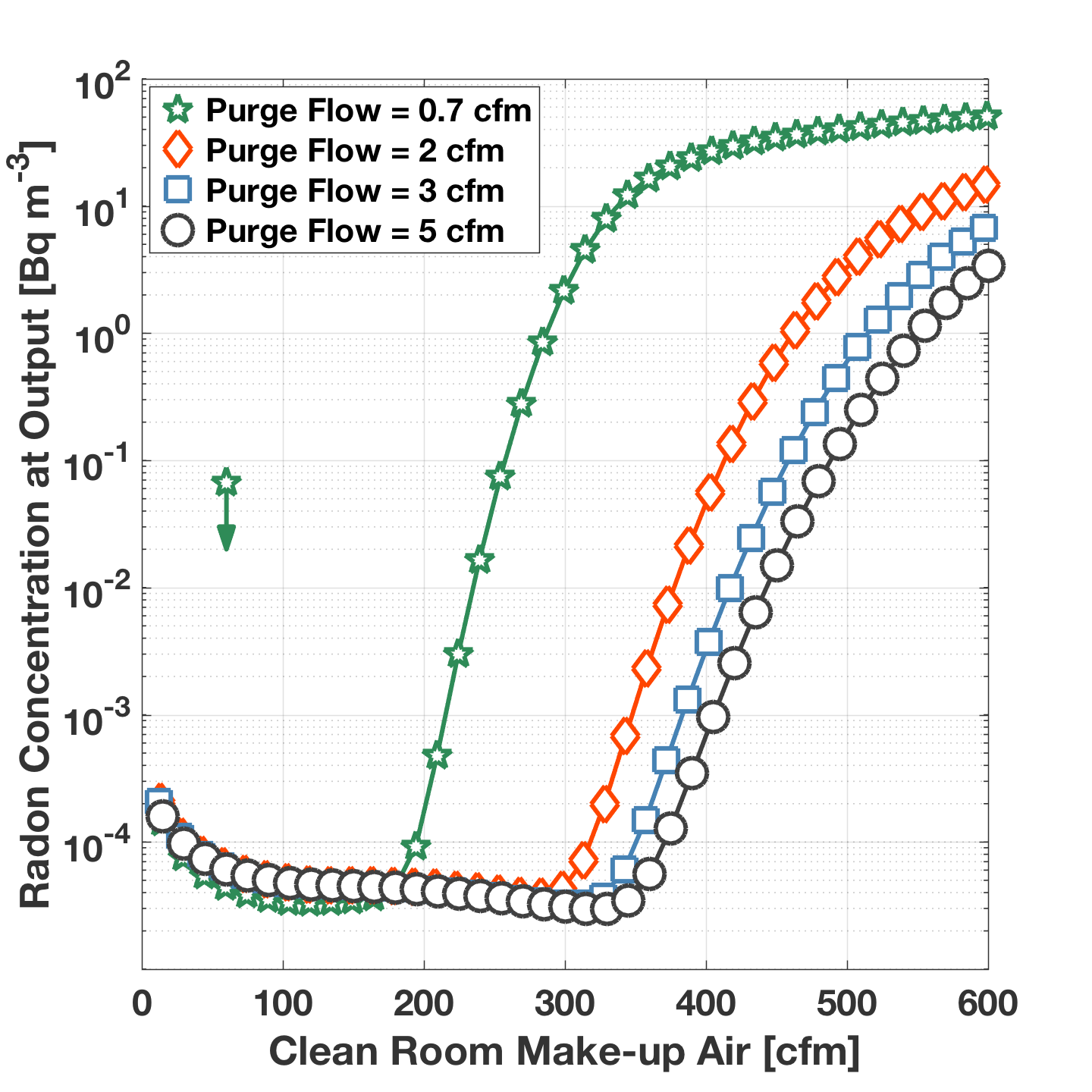}
	\end{tabular}
	\caption{\textit{Left}: A measured break-through curve (vertical error bars) at 90\,cfm behind a simulated break-through curve (solid) with a reduced $\chi^2$ of 0.98 and p-value of 0.22. This fit suggests $D=32\,$cm$^2$\,hour$^{-1}$ and $A_\circ=1.17$ for a breakthrough time $t_\mathrm{BT}=9.26\,$hours. The dashed line represents the radon concentration at the input, shifted by the break-through time, accounting for radon decay. \textit{Right}: Predicted radon concentration at the VSA system output versus clean room make-up air flow for various purge flows (curves), with a 100\,Bq\,m$^{-3}$ input radon concentration, compared to the measured upper limit of 0.067\,Bq\,m$^{-3}$ shown in Fig.~\ref{CR}. The star-curve represents the system as configured in February 2016 while the other curves represent the current system. The simulation shows that purge flow above 5 cfm does not further reduce the radon concentration. Radon emanating from the activated carbon dominates the radon concentration at the output for flows $<$200\,cfm. at low make-up air flow ($<$10\,cfm) and determines the lower limit on radon concentration. The contribution of radon emanating from the carbon decreases with increasing make-up air flow, until, at higher flows, radon begins breaking-through the columns due to only partial regeneration.}
	\label{fig:BT_Curve}
\end{figure}
For the simulation to make accurate predictions, internal parameters must be inferred from measurements. These parameters are the diffusion coefficient $D$ and the radon adsorption factor $A_\circ$, both of which depend on details of the carbon and its packing.

The internal parameters are determined with so-called break-through curves (Fig.~\ref{fig:BT_Curve}, \textit{Left}). A break-through curve is produced when high radon air flows through a single, regenerated carbon column. At first, the air exiting the column was low radon. Radon adsorbs and desorbs from the carbon surfaces as it moves through a carbon column and diffuses over time according to Eq.~\ref{dist}. Eventually, radon begins ``breaking through" the column until the output of the column is similar to the input delayed by the breakthrough time but slightly decayed and diffused. The diffusion coefficient determines the spread of the radon distribution and $A_\circ$ determines its translation (modifying the radon velocity). Figure~\ref{fig:BT_Curve} (\textit{Left}) shows that the simulation agrees well with the data, although small systematic shifts suggest an even more accurate model may be possible.

The second characterization is from the equilibrium radon concentration at the VSA system output. This comparison is important because it examines not only the forward flow of the simulation but also regeneration and slow fill. Regeneration and slow-fill components of the simulation may be examined closely. This form of characterization is currently ongoing. 

Once the simulation has been characterized, many VSA system operating parameters may be explored. Figure~\ref{fig:BT_Curve} (\textit{Right}) shows the simulated radon concentration at the VSA system output versus clean room make-up air flow for several values of the purge flow. The star-curve represents the VSA system configuration during February 2016. The other curves represent radon concentration at output for the current system configured for 40\,min of regeneration, 15\,min of slow-fill, and purge flows of 2, 3, or 5\,cfm. The simulation indicates that the VSA system should achieve radon reduction $>$100$\times$ larger than is typically achieved with (more expensive) continuous flow systems.

\section{ACKNOWLEDGMENTS}
This work was supported in part by the National Science Foundation (Grant No. PHY-1506033 and EEC-1461190) and the Pacific Northwest National Laboratory (PNNL), operated by Battelle for the U.S. Department of Energy (DOE) under Contract Nos. DE-AC05-76RL01830.



\bibliographystyle{aipnum-cp}
\bibliography{LRT2017VSA}

\end{document}